\documentclass[twoside,twocolumn,9pt]{article}
\usepackage{extsizes}
\usepackage[super,sort&compress,comma]{natbib} 
\usepackage[version=3]{mhchem}
\usepackage[left=1.5cm, right=1.5cm, top=1.785cm, bottom=2.0cm]{geometry}
\usepackage{balance}
\usepackage{mathptmx}
\usepackage{sectsty}
\usepackage{graphicx} 
\usepackage{lastpage}
\usepackage[format=plain,justification=justified,singlelinecheck=false,font={stretch=1.125,small,sf},labelfont=bf,labelsep=space]{caption}
\usepackage{float}
\usepackage{fancyhdr}
\usepackage{fnpos}
\usepackage[english]{babel}
\addto{\captionsenglish}{%
  
}
\usepackage{array}
\usepackage{droidsans}
\usepackage{charter}
\usepackage[T1]{fontenc}
\usepackage[usenames,dvipsnames]{xcolor}
\usepackage{setspace}
\usepackage[compact]{titlesec}
\usepackage{hyperref}
\usepackage{epstopdf}%This line makes .eps figures into .pdf - please comment out if not required.

\definecolor{cream}{RGB}{222,217,201}

\begin{document}

\pagestyle{fancy}
\thispagestyle{plain}
\fancypagestyle{plain}{
%%%HEADER%%%
\renewcommand{\headrulewidth}{0pt}
}
%%%END OF HEADER%%%

%%%PAGE SETUP - Please do not change any commands within this section%%%
\makeFNbottom
\makeatletter
\renewcommand\LARGE{\@setfontsize\LARGE{15pt}{17}}
\renewcommand\Large{\@setfontsize\Large{12pt}{14}}
\renewcommand\large{\@setfontsize\large{10pt}{12}}
\renewcommand\footnotesize{\@setfontsize\footnotesize{7pt}{10}}
\makeatother

\renewcommand{\thefootnote}{\fnsymbol{footnote}}
\renewcommand\footnoterule{\vspace*{1pt}% 
\color{cream}\hrule width 3.5in height 0.4pt \color{black}\vspace*{5pt}} 
\setcounter{secnumdepth}{5}

\makeatletter 
\renewcommand\@biblabel[1]{#1}            
\renewcommand\@makefntext[1]% 
{\noindent\makebox[0pt][r]{\@thefnmark\,}#1}
\makeatother 
\renewcommand{\figurename}{\small{Fig.}~}
\sectionfont{\sffamily\Large}
\subsectionfont{\normalsize}
\subsubsectionfont{\bf}
\setstretch{1.125} %In particular, please do not alter this line.
\setlength{\skip\footins}{0.8cm}
\setlength{\footnotesep}{0.25cm}
\setlength{\jot}{10pt}
\titlespacing*{\section}{0pt}{4pt}{4pt}
\titlespacing*{\subsection}{0pt}{15pt}{1pt}
%%%END OF PAGE SETUP%%%

%%%FOOTER%%%
\fancyfoot{}
\fancyfoot[LO,RE]{\vspace{-7.1pt}\includegraphics[height=9pt]{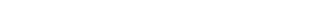}}
\fancyfoot[CO]{\vspace{-7.1pt}\hspace{13.2cm}\includegraphics{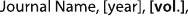}}
\fancyfoot[CE]{\vspace{-7.2pt}\hspace{-14.2cm}\includegraphics{head_foot/RF}}
\fancyfoot[RO]{\footnotesize{\sffamily{1--\pageref{LastPage} ~\textbar  \hspace{2pt}\thepage}}}
\fancyfoot[LE]{\footnotesize{\sffamily{\thepage~\textbar\hspace{3.45cm} 1--\pageref{LastPage}}}}
\fancyhead{}
\renewcommand{\headrulewidth}{0pt} 
\renewcommand{\footrulewidth}{0pt}
\setlength{\arrayrulewidth}{1pt}
\setlength{\columnsep}{6.5mm}
\setlength\bibsep{1pt}
%%%END OF FOOTER%%%

%%%FIGURE SETUP - please do not change any commands within this section%%%
\makeatletter 
\newlength{\figrulesep} 
\setlength{\figrulesep}{0.5\textfloatsep} 

\newcommand{\topfigrule}{\vspace*{-1pt}% 
\noindent{\color{cream}\rule[-\figrulesep]{\columnwidth}{1.5pt}} }

\newcommand{\botfigrule}{\vspace*{-2pt}% 
\noindent{\color{cream}\rule[\figrulesep]{\columnwidth}{1.5pt}} }

\newcommand{\dblfigrule}{\vspace*{-1pt}% 
\noindent{\color{cream}\rule[-\figrulesep]{\textwidth}{1.5pt}} }

\makeatother
%%%END OF FIGURE SETUP%%%

%%%TITLE, AUTHORS AND ABSTRACT%%%
\twocolumn[
  \begin{@twocolumnfalse}
{\includegraphics[height=30pt]{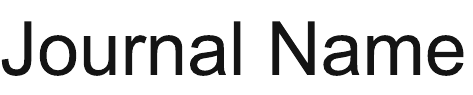}\hfill\raisebox{0pt}[0pt][0pt]{\includegraphics[height=55pt]{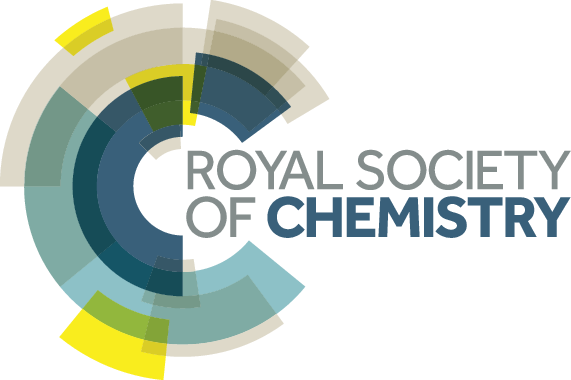}}\\[1ex]
\includegraphics[width=18.5cm]{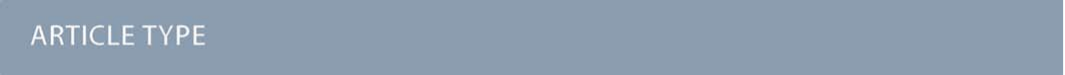}}\par
\vspace{1em}
\sffamily
\begin{tabular}{m{4.5cm} p{13.5cm} }

\includegraphics{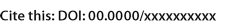} & \noindent\LARGE{\textbf{Artificial Intelligence Approaches for Anti-Addiction Drug Discovery}} \\%Article title goes here instead of the text "This is the title"
\vspace{0.3cm} & \vspace{0.3cm} \\

 & \noindent\large{Dong Chen\textit{$^{a}$}, Jian Jiang,\textit{$^{b,a}$}, Zhe Su\textit{$^{a}$}, and Guo-Wei Wei$^{\ast}$\textit{$^{a,c,d}$}} \\%Author names go here instead of "Full name", etc.
 
\includegraphics{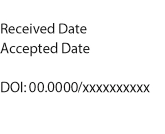} & \noindent\normalsize{Drug addiction is a complex and pervasive global challenge that continues to pose significant public health concerns. Traditional approaches to anti-addiction drug discovery have struggled to deliver effective therapeutics, facing high attrition rates, long development timelines, and inefficiencies in processing large-scale data. Artificial intelligence (AI) has emerged as a transformative solution to address these issues. Using advanced algorithms, AI is revolutionizing drug discovery by enhancing the speed and precision of key processes. This review explores the transformative role of AI in the pipeline for anti-addiction drug discovery, including data collection, target identification, and compound optimization. By highlighting the potential of AI to overcome traditional barriers, this review systematically examines how AI addresses critical gaps in anti-addiction research, emphasizing its potential to revolutionize drug discovery and development, overcome challenges, and advance more effective therapeutic strategies.} \\%The abstrast goes here instead of the text "The abstract should be..."

\end{tabular}

 \end{@twocolumnfalse} \vspace{0.6cm}

  ]
%%%END OF TITLE, AUTHORS AND ABSTRACT%%%

%%%FONT SETUP - please do not change any commands within this section
\renewcommand*\rmdefault{bch}\normalfont\upshape
\rmfamily
\section*{}
\vspace{-1cm}

%%%FOOTNOTES%%%
\footnotetext{\textit{$^{a}$~Department of Mathematics, Michigan State University, MI, 48824, USA}}
\footnotetext{\textit{$^{b}$~Research Center of Nonlinear Science, School of Mathematical and Physical Sciences, Wuhan Textile University, Wuhan, 430200, P.R. China}}
\footnotetext{\textit{$^{c}$~Department of Electrical and Computer Engineering, Michigan State University, MI 48824, USA}}
\footnotetext{\textit{$^{d}$~Department of Biochemistry and Molecular Biology, Michigan State University, MI 48824, USA; E-mail: weig@msu.edu}}

%Please use \dag to cite the ESI in the main text of the article.
%If you article does not have ESI please remove the the \dag symbol from the title and the footnotetext below.
% \footnotetext{\dag~Supplementary Information available: [details of any supplementary information available should be included here]. See DOI: 00.0000/00000000.}

% \footnotetext{\ddag~Additional footnotes to the title and authors can be included \textit{e.g.}\ `Present address:' or `These authors contributed equally to this work' as above using the symbols: \ddag, \textsection, and \P. Please place the appropriate symbol next to the author's name and include a \texttt{\textbackslash footnotetext} entry in the the correct place in the list.}

%%%END OF FOOTNOTES%%%

%%%MAIN TEXT%%%%
\section{Introduction}\label{section:introduction}

Drug addiction is a chronic relapsing disease that affects millions of people globally and poses significant social, economic, and health challenges\cite{rehman2022impacts,peacock2018global,hansford2019world,f2011understanding}. It is a syndromic condition characterized by forcing the body to compulsively engage in rewarding stimuli regardless of adverse consequences. The underlying biology involves complex neurochemical pathways such as the dopaminergic reward system \cite{arias2010dopaminergic}, glutamatergic circuits that govern synaptic plasticity \cite{kalivas2009glutamate}, and the GABAergic system (GABA) responsible for inhibitory control \cite{feil2010addiction}. The GABAergic system is responsible for inhibitory control, and so on. These systems interact with genetic predispositions, epigenetic modifications, and environmental factors to form a complex network of mechanisms that drive addiction \cite{volkow2011addiction}. Genetic studies have identified many polymorphisms associated with susceptibility, and advances in neuroimaging have revealed brain regions associated with craving and relapse \cite{mackey2016genetic}. However, despite this current understanding, understanding the molecular basis of addiction remains a challenge.

Addiction treatment is a pressing global problem that requires innovative strategies for effective intervention. The discovery of anti-addiction drugs plays a key role in mitigating the effects of addiction. Efforts in this field have focused on the development of therapeutic drugs that modify addictive behaviors, reduce withdrawal symptoms, and prevent relapse \cite{koob2009development}. However, the discovery process for addiction therapeutics faces many obstacles, including the intricate neurobiology of substance use disorders (SUDs), high attrition rates during clinical trials, and lengthy timeframes associated with traditional approaches. Empirical screening and receptor-based drug design have yielded important insights and treatments \cite{johnson2011opportunities}, but there are limitations in addressing the complexity of addiction mechanisms\cite{suggitt200550,lavecchia2013virtual,tanrikulu2008pseudoreceptor}. Challenges posed by traditional approaches have highlighted the need for advanced methods. Thanks to the development of Artificial intelligence (AI) technology and the accumulation of data related to drug design, data-driven-based approaches provide efficient insights for addiction research.

AI has become an important tool in anti-addiction research, providing innovative solutions to the limitations of traditional methodologies \cite{mak2024artificial}. AI has dramatically accelerated the drug discovery process by using machine learning algorithms to analyze large chemical databases, identify potential therapeutic targets, and predict drug efficacy \cite{mak2024artificial,vora2023artificial,suva2024artificial}. These AI-driven tools optimize the screening of billions of compounds, significantly reducing the time and cost associated with traditional drug development \cite{murugan2022artificial}. In addition, AI enables the integration of large datasets of protein structure and molecular information to design novel therapeutic drugs for addictive conditions, while also providing breakthrough pathways for treating complex addictions such as opioid dependence.

Beyond drug discovery, AI is transforming addiction treatment through personalized treatment strategies. Advanced algorithms can analyze genetic information, medical history, and behavioral patterns to create personalized treatment plans \cite{chhetri2023machine}. By incorporating factors such as genetic predisposition, social environment, and mental health status, AI-driven models can predict the risk of relapse, leading to proactive and timely interventions \cite{beaulieu2021artificial}. Another major advantage of AI is the ability to process and analyze large-scale data from various sources, including electronic health records and wearable devices \cite{chen2021applications}. Utilizing natural language processing and machine learning techniques, AI can generate real-world evidence to inform treatment decisions and improve outcomes \cite{yang2022large}. These insights provide generalizable measures of effectiveness, address critical gaps in addiction research, and ensure better integration of data into clinical practice. Artificial intelligence has become an indispensable tool for overcoming the challenges of traditional anti-addiction drug discovery methods, driving progress in discovering and developing innovative treatment solutions.

This review explores the role that AI-driven approaches play in the anti-addiction drug research process and how it is reshaping the field, as well as how it is addressing long-standing gaps in addiction therapies \cite{feng2022machine,mak2024artificial,pushpakom2019drug,mcnair2023artificial,koob2009development}. The scope of this review covers the key stages of the drug discovery process, including data collection, target identification, and lead compound discovery and optimization. We will look at each of these steps from the perspective of artificial intelligence applications, emphasizing how machine learning models and computational tools can improve the efficiency and accuracy of these processes. Data collection for anti-addiction research involves integrating various datasets such as genetic information, neurobiological data, and compound databases to create a comprehensive analytical resource. Artificial intelligence-driven approaches provide powerful tools for processing, coordinating, analyzing, and interpreting large-scale datasets. In the area of target identification, AI has accelerated the discovery of molecular and neurobiological targets associated with addiction, facilitating the identification of pathways and receptors that can be modulated to mitigate addictive behaviors. Similarly, in the area of lead compound discovery and optimization, AI-driven technologies simplify the screening and design of potential drug candidates, providing solutions that reduce the high attrition rates and lengthy timelines traditionally associated with this phase.

By systematically addressing these stages, this review aims to provide a comprehensive understanding of how AI-based models are revolutionizing anti-addiction drug discovery, showcasing their potential to overcome long-standing challenges and advance the development of effective therapeutics.

\section{Data Collection}

% Data Collection
%  3. **Datasets for AI in Anti-Addiction Drug Discovery**
% - **Molecular Datasets**:
%   - Chemical compound libraries (e.g., ChEMBL, PubChem, ZINC database).
%   - Bioactivity databases specific to addiction targets.
% - **Biological Data**:
%   - Proteomics and genomics datasets for target identification.
%   - Pharmacokinetics and pharmacodynamics data.
% - **Behavioral and Clinical Data**: （optional)
%   - Data from animal models and clinical trials of addiction therapies.
%   - Patient outcomes related to existing treatments.
% - Challenges: data heterogeneity, noise, and limited datasets specific to addiction.

% ### Key Databases and Tools
% - **PubChem and ChEMBL**: Contain compound and bioactivity data for training AI models.
% - **OpenTargets**: Provides insights into drug-target-disease associations.
% - **Addiction-Specific Databases**: Genomic data from addiction studies (e.g., GWAS Catalog) can be integrated into AI workflows.

In AI-driven drug discovery, datasets are fundamental as they directly impact the performance and accuracy of AI models. A robust and integrated dataset ecosystem is crucial for developing reliable models to tackle complex challenges, particularly in areas like anti-addiction therapeutics.

\subsection{Molecular Datasets}  
Molecular datasets are invaluable for drug discovery, providing chemical and pharmacological information essential for identifying candidate molecules. General purpose experimental databases such as ChEMBL\cite{gaulton2012chembl}, PubChem\cite{kim2016pubchem}, and DrugBank\cite{wishart2018drugbank} provide comprehensive data on drug-like molecules, including their structures, properties, and interactions with biological targets. These databases are fundamental for building predictive models, screening drug candidates, and exploring existing therapies for repurposing. Among these, DrugBank is particularly notable for anti-addiction drug discovery, as it integrates data on FDA-approved drugs, experimental therapeutics, and pharmacological properties\cite{wishart2018drugbank}. 

For addiction-focused studies, specialized molecular datasets provide deeper insight into receptor-ligand interactions and toxicity profiles. BindingDB, for instance, offers protein-ligand binding affinity data critical for understanding receptor-ligand dynamics in addiction-related pathways\cite{liu2007bindingdb}. To address safety considerations, toxicity-focused datasets such as Tox21 provide chemical toxicity information, aiding researchers in designing safer therapeutic agents\cite{richard2020tox21}. Complementing these experimental resources are computational datasets such as the ZINC Database\cite{irwin2005zinc}, which contains millions of virtual molecules suitable for virtual screening, allowing efficient exploration of chemical spaces and cost-effective identification of potential therapeutic compounds. Advancements in machine learning have further enhanced the utility of molecular datasets in drug discovery. For example, the MolData dataset compiles extensive PubChem drug screening results, facilitating molecular machine learning applications aimed at improving drug discovery processes\cite{keshavarzi2022moldata}. By providing a structured compilation of bioassay data, MolData enables the development of predictive models that can identify potential therapeutic compounds and repurposing opportunities across various diseases, including addiction.

Addressing specific addiction-related mechanisms and safety challenges requires targeted molecular datasets. For psychostimulant drugs such as cocaine, the primary therapeutic strategy involves inhibiting dopamine reuptake by modulating the dopamine transporter (DAT). The DAT dataset is crucial in this context, offering insight into DAT interactions, a key element in addiction-related pathways\cite{zhu2023tidal}. By elucidating the biochemical mechanisms of dopamine reuptake, this dataset supports the design of compounds that precisely modulate DAT activity, targeting addiction at its core. However, the development of DAT inhibitors necessitates careful evaluation of off-target effects, particularly the potential blockade of the human Ether-\`{a}-go-go-Related Gene (hERG) potassium channel, which could lead to severe ventricular arrhythmias. The hERG dataset addresses this critical safety concern by assessing the cardiotoxic potential of candidate molecules\cite{zhu2023tidal,zhang2022hergspred}. Together, the DAT and hERG datasets enable researchers to design addiction therapies that balance efficacy and safety, providing a targeted approach with rigorous safety profiling\cite{lee2021toward}.

\subsection{Biological Datasets} 
Biological datasets are key resources for identifying drug targets and understanding the complex mechanisms of addiction. These resources provide a solid foundation for anti-addiction research, enabling researchers to discover therapeutic targets and develop safe and effective interventions. Gene expression datasets, such as Gene Expression Omnibus (GEO) \cite{clough2016gene}, Encyclopedia of DNA Elements (ENCODE) \cite{encode2011user}, and ArrayExpress \cite{parkinson2007arrayexpress}, are of particular value for studying addiction and anti-addiction drug discovery. For example, GEO contains a curated dataset of gene expression profiles, providing insights into gene and pathway imbalances in addiction contexts. ENCODE extends these features by providing a comprehensive catalog of functional elements of the human genome, including transcriptional and epigenetic data, which can be used to gain insight into gene regulatory mechanisms associated with addiction and discover target genes. In addition, ArrayExpress stores data from high-throughput functional genomics experiments, enabling researchers to analyze transcript changes associated with addiction.

In addition, protein-protein interaction (PPI) databases such as STRING\cite{mering2003string}, BioGRID\cite{stark2006biogrid}, and IntAct\cite{hermjakob2004intact} have also enriched the tools available in the drug discovery process by providing a framework for studying addiction at a systems level. Among them, the STRING dataset integrates experimental data and predicted data to map interaction networks, allowing researchers to identify proteins involved in addiction-related signaling pathways, such as proteins in dopaminergic and serotonergic systems found in \citet{mering2003string}'s studies. BioGRID provides experimentally validated data to discover high-confidence targets, while IntAct provides curated interaction profiles that can be used to prioritize addiction-related proteins\cite{stark2006biogrid}. Together, these PPI datasets support pathway enrichment analysis and the identification of key proteins in addiction-related networks.

Complementing these datasets, the Human Protein Atlas links protein expression and localization data to specific tissues, including regions of the brain affected by addiction. This enables the identification of tissue-specific drug targets, minimizing off-target effects and improving treatment precision. The Comparative Toxicogenomics Database (CTD) links genetic data to chemical exposures, revealing how environmental agents affect addiction pathways and guiding therapeutic development\cite{davis2023comparative}. The Human Reference Interactome (HuRI) also provides a high-quality PPI network, highlighting central proteins (hubs) as major drug targets\cite{luck2020reference}.

By integrating these datasets, researchers can prioritize targets, perform pathway enrichment analysis, and conduct research and development for anti-addiction drugs. In addition, a collection of the molecular and biological datasets and their references is provided in Table~\ref{tbl:datasets}.

% KEGG (Kyoto Encyclopedia of Genes and Genomes)
% Description: A database integrating genomic, chemical, and systemic functional information.
% Use Cases: Pathway analysis, target identification, and compound screening.

\begin{table}[!ht]
\small
  \centering
  \caption{Summary of molecular and biological datasets}
  \label{tbl:datasets}
  \begin{tabular*}{0.48\textwidth}{@{\extracolsep{\fill}}cc}
    % \begin{tabular}{cccc}
      \hline
      % Molecular Datasets &  Refs                        & Biological Datasets & Refs    \\ \hline 
      Molecular Datasets                        & Biological Datasets  \\ \hline 
      PubChem      \cite{kim2016pubchem}        & STRING              \cite{mering2003string} \\
      ChEMBL       \cite{gaulton2012chembl}     & BioGRID             \cite{stark2006biogrid} \\
      DrugBank     \cite{wishart2018drugbank}   & IntAct              \cite{hermjakob2004intact} \\
      BindingDB    \cite{liu2007bindingdb}      & \begin{tabular}[c]{@{}c@{}}Comparative Toxicogenomics Database\\(CTD)\end{tabular} \cite{davis2023comparative} \\
      ZINC         \cite{irwin2005zinc}         & \begin{tabular}[c]{@{}c@{}}Human Reference Interactome\\(HuRI)\end{tabular}\cite{luck2020reference} \\
      MolData      \cite{keshavarzi2022moldata} & \begin{tabular}[c]{@{}c@{}}Gene Expression Omnibus\\(GEO)\end{tabular} \cite{clough2016gene} \\
      Tox21        \cite{huang2016modelling}    & \begin{tabular}[c]{@{}c@{}}Encyclopedia of DNA Elements\\(ENCODE)\end{tabular}  \cite{encode2011user} \\
      GDB-17       \cite{ruddigkeit2012enumeration}& ArrayExpress \cite{parkinson2007arrayexpress}    \\
      hERG         \cite{zhang2022hergspred}    & \begin{tabular}[c]{@{}c@{}}Kyoto Encyclopedia of Genes and Genomes\\(KEGG)\end{tabular} \cite{kanehisa2000kegg} \\
      DAT          \cite{lee2021toward}         & \begin{tabular}[c]{@{}c@{}}Protein Data Bank\\(PDB)\end{tabular} \cite{berman2002protein} \\
                                                & Open Targets  \cite{koscielny2017open} \\
      \hline
      \end{tabular*}
  \end{table}

\section{Target Identification}
Target identification is a critical step in drug discovery, and in addiction research, this step involves pinpointing biological components that modulate addictive behaviors or are affected by substance use disorders, often involving neurological \cite{agrawal2012genetics,popescu2021understanding}, genetic \cite{popescu2021understanding}, and behavioral factors \cite{grant2010introduction}. The complexity of addiction, stemming from neurochemical, genetic, and environmental interactions that influence brain reward circuits and neurotransmitters such as dopamine and glutamate, complicates the discovery of effective therapeutic targets. Traditional approaches have difficulty navigating these intricacies, hampering progress in drug discovery against addiction \cite{white2020human} due to overlapping pathways \cite{griffiths2008biopsychosocial}, limited high-quality data \cite{gangwal2024current}, and ethical restrictions on human studies. The emergence of AI can revolutionize the process of target identification by analyzing large data sets, detecting patterns, and accurately predicting treatment targets \cite{cresta2022practical1,cresta2022practical2}. By leveraging machine learning, researchers can integrate multi-omics data, such as genomics, proteomics, and metabolomics, to reveal hidden pathways associated with addiction and provide key insights into the interaction between genetics and the environment in addiction biology \cite{singh2023unveiling,guan2022integrative,singh2023integrative}. This section focuses on the application of AI in target identification for anti-addiction research.
  
  \subsection{AI for Genomic Data and Multi-Omics Integration}
  Deep learning models have shown remarkable potential in the analysis of genomic and transcriptomic data \cite{koumakis2020deep,yue2023deep}. These models are particularly adept at processing high-throughput sequencing data, enabling the identification of genes and pathways associated with various traits and conditions, including addiction. Among these models, Convolutional Neural Networks (CNNs) excel in capturing both local and global patterns within genomic data, making them highly effective in identifying functional genetic variants and unraveling intricate biological structures. For example, CNNs have been successfully applied to predict phenotypes from single nucleotide polymorphisms (SNPs), thus uncovering genetic markers associated with specific traits \cite{liu2019phenotype}. This capability underscores their utility in the identification of potential therapeutic targets.

  \begin{figure*}[!ht]
     \centering
      \includegraphics[width=\textwidth]{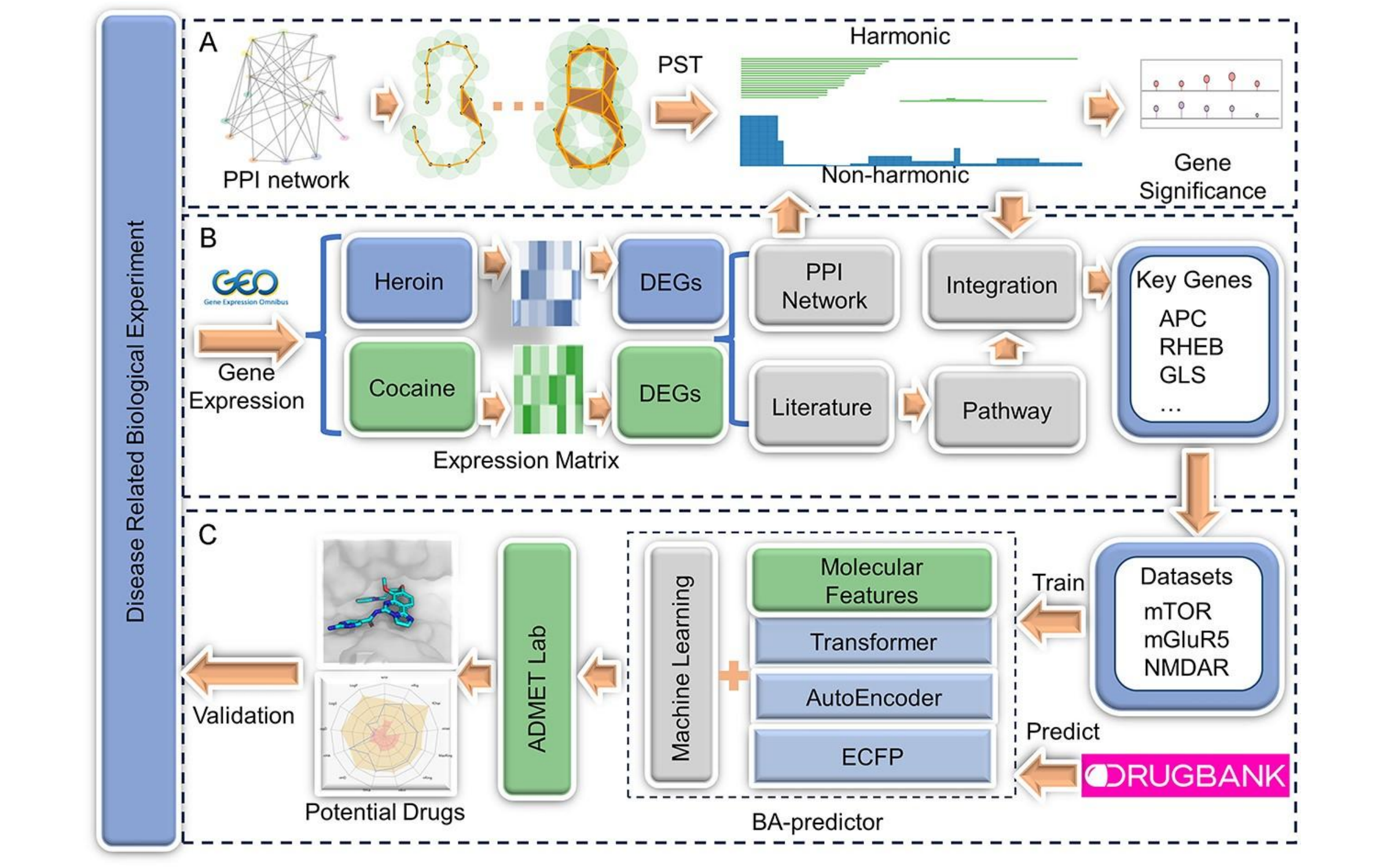}
      % \captionsetup{width=16cm}
      \caption{
      {\bf a} PST-Based Topological Differentiation Analysis: The persistent spectral graph was used to identify key nodes in the PPI network.  
      {\bf b} Gene Analysis: Opioid- and cocaine-related transcriptomic data from GEO were analyzed to identify key genes through PPI networks, validated via literature and pathway analysis.  
      {\bf c} Drug Repurposing: Machine learning models predicted DrugBank compound affinities for addiction-related targets (mTOR, mGluR5, NMDAR), with ADMET analysis identifying potential treatments.
      Reproduced with permission from ref \cite{du2024multiscale}. Copyright 2024 Oxford University Press.
      }
      \label{fig:genomic}
  \end{figure*}
  
  In genomic selection, CNN-based frameworks have outperformed traditional statistical methods by accurately predicting quantitative phenotypes without requiring imputation. They efficiently identify significant genotype markers and provide detailed information on trait contributions using techniques such as saliency mapping. In addition, CNNs have been used to predict antibiotic resistance phenotypes in Mycobacterium tuberculosis, where they identified novel genomic loci associated with resistance \cite{green2022convolutional}. These applications demonstrate CNNs' ability to generate biologically meaningful interpretations and actionable insights, directly informing the discovery of molecular targets relevant to addiction mechanisms. Recurrent neural networks (RNNs) also excel in handling sequence data, which makes them well-suited to analyze DNA sequences \cite{yue2023deep}. Other language models, such as Long Short-Term Memory networks (LSTMs) and Transformers, have been applied to processing long genomic sequences \cite{nguyen2024hyenadna}, identifying patterns in gene expression data \cite{choi2023transformer}, and predicting gene function based on sequence information \cite{ji2021dnabert}. These models have demonstrated promising performance in target identification within the drug discovery process, including addiction research.
  
  Genomic data serves as the cornerstone of multi-omics integration, providing the initial genetic information upon which other omics layers, such as transcriptomics, proteomics, and metabolomics, build to offer a systems-level understanding of biology. AI-driven multi-omics integration has emerged as a transformative approach in addiction research and drug discovery, combining data such as various biomolecular levels—DNA, RNA, proteins, metabolites, and epigenetic marks to provide a comprehensive understanding of addiction mechanisms \cite{ivanisevic2023multi}. By integrating these diverse datasets, researchers can uncover addiction-specific biomarkers and prioritize potential drug targets, paving the way for personalized treatments and more effective therapies. For example, \citet{du2024multiscale} developed a versatile framework for drug repurposing in opioid and cocaine addiction, integrating transcriptomic analysis, topological data methods, and machine learning (as shown in Figure~\ref{fig:genomic}). They identified three key targets—mTOR, mGluR5, and NMDAR—and validated potential drugs using predictive models.

  % ### 2. **Knowledge Graphs for Drug-Target Interaction to identify the protein target**
  
  \subsection{Knowledge Graphs for Drug-Target Interaction}
  Knowledge graphs (KGs) have become powerful tools in drug discovery, particularly in predicting drug-target interactions (DTIs) and identifying potential protein targets, including applications in addiction research. By integrating diverse biological data, such as gene expression, protein-protein interactions, clinical findings, and other biological entities, these graphs create comprehensive networks that illuminate relationships between drugs, targets, and diseases \cite{zhou2024tarkg,liu2024probabilistic}.  Graph neural network (GNN) based techniques, such as DeepWalk \cite{zhao2021novel} and MLGANN \cite{zhao2021novel}, enable the identification of network hubs and patterns that highlight potential drug targets. In addition, cutting-edge models such as DTKGIN \cite{luo2024dtkgin}, GraphDTA \cite{nguyen2021graphdta}, and KGE-UNIT \cite{zhang2024kge} improve DTI prediction accuracy by learning the feature representations of biological entities, even in the absence of prior experimental data.

  \begin{figure*}[!ht]
      \centering
      \includegraphics[width=0.9\textwidth]{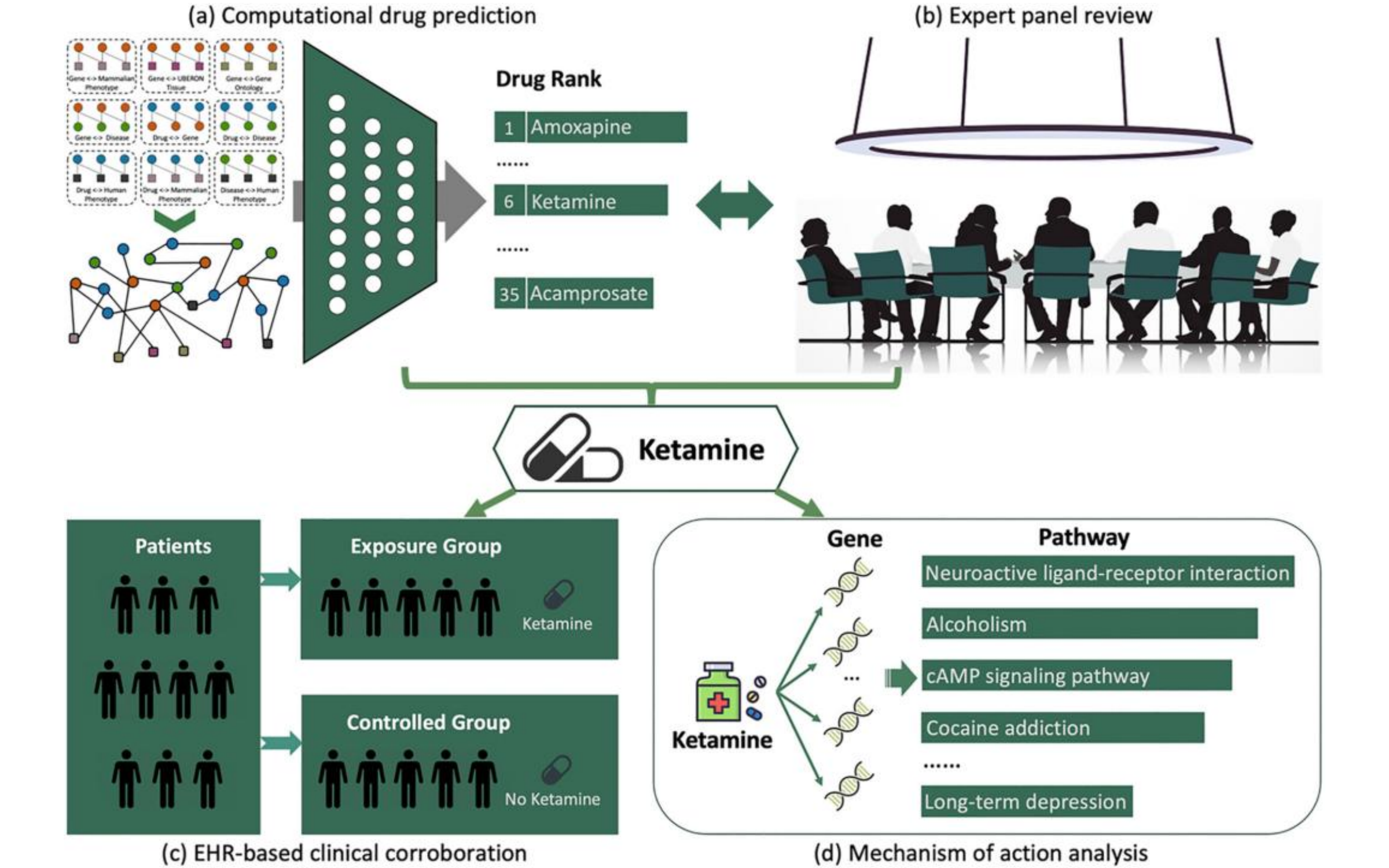}
      % \captionsetup{width=16cm}
      \caption{Overview of the drug repurposing pipeline for cocaine use disorder. 
          {\bf a} A knowledge graph-based drug discovery system integrating multi-type interactions from diverse biomedical databases was utilized to rank potential candidate drugs for CUD treatment.  
          {\bf b} The CTN-0114 advisory committee evaluated the top-ranked candidates and selected ketamine for clinical investigation.  
          {\bf c} Insights from electronic health records provided clinical evidence supporting ketamine's potential effectiveness for CUD treatment.  
          {\bf d} Genetic and functional analyses revealed that ketamine directly interacts with multiple CUD-associated genes and pathways.
      Reproduced with permission from ref \cite{gao2023repurposing}. Copyright 2023 Wiley.
      }
      \label{fig:knowledge_graph}
  \end{figure*}

  These methodologies have transformative potential in addiction research and drug discovery, advancing target identification, drug repurposing, mechanism understanding, and prediction accuracy. They integrate diverse biological data, enabling computational approaches like Progeni \cite{liu2024probabilistic}, which uses a GNN-based model to predict biologically relevant targets with superior accuracy, validated in studies on melanoma and colorectal cancer. For drug repurposing, KGs like OREGANO\cite{boudin2023oregano} compile comprehensive datasets, facilitating hypothesis generation and community collaboration. Mechanism understanding is enhanced by tailored KGs such as TarKG\cite{zhou2024tarkg}, which integrates vast biomedical and traditional medicine data to predict disease-target relationships, as demonstrated in Alzheimer's research. \citet{gao2023repurposing} identifies ketamine as a promising repurposed drug for cocaine use disorder through an integrated strategy, as shown in Figure~\ref{fig:knowledge_graph}, which combines knowledge graph-based prediction, expert review, clinical data, and mechanism analysis. In addition, advanced models such as Multi-Layer Graph Attention Neural Network (MLGANN)\cite{lu2024multi} improve drug-target interaction predictions by combining multi-source data and network heterogeneity, offering unparalleled precision. Using these knowledge graph-based approaches, researchers can accelerate the discovery of new targets and potential treatments for addiction, ultimately improving the drug development process in this critical area of research.

  % ### 3. **AI-Powered Inverse Virtual Screening for target identification in anti-addiction research**
  \subsection{Inverse Virtual Screening}
  AI-driven reverse virtual screening has emerged as an effective tool for target identification in anti-addiction research, which provides a cost-effective approach to drug discovery. Using artificial intelligence, this method is able to screen large protein libraries and predict drug-target interactions, thereby helping to identify specific molecular targets and therapeutic candidates. For example, \citet{schottlender2022drugs} introduces a reverse virtual screening strategy for predicting potential targets of antimicrobial compounds and reduces putative targets significantly, and aims to enhance antimicrobial drug discovery via integration into the Target Pathogen database. Through advanced AI technology, protein library screening can quickly identify potential targets for addiction treatment drugs. This approach is particularly valuable when there are few known addiction-related protein targets because it can discover similar proteins that can also serve as viable targets \cite{zhou2024artificial}. In addition, artificial intelligence frameworks excel in predicting interactions (such as binding affinity) between known drugs and potential protein targets \cite{chen2024multiscale}, thereby supporting drug repurposing efforts. These methods enable researchers to explore the molecular targets of anti-addiction drugs and accelerate the discovery of new treatment options \cite{song2024drug,abbasi2023drug}.

  In addition, natural language processing (NLP) models (e.g., ESM\cite{rives2021biological} for protein sequences) and SMILES-based models (e.g., transformer-CPZ\cite{chen2021extracting}) provide more opportunities for sequence-based screening. Their ability to handle complex biological sequences makes them promising tools for anti-addiction research \cite{feng2023virtual,feng2022machine}. Generative adversarial networks (GANs), including GAN-WGCNA, have been used to analyze gene expression data to identify key regulators in addiction pathways. For example, studies on cocaine self-administration using GAN-WGCNA revealed key genes associated with addictive behavior, such as Alcam and Celf4 \cite{kim2024gan}. Similarly, variational autoencoders (VAEs) are effective in revealing nonlinear underlying patterns in metabolomics data, revealing addiction-related processes, and highlighting potential therapeutic targets \cite{gomari2022variational}.

  This AI-driven approach offers significant advantages, including the reduction of expensive and time-consuming experimental screening and the discovery of novel drug-target interactions relevant to addiction. By integrating these advanced methodologies, researchers can efficiently identify and validate therapeutic targets, ultimately accelerating the development of effective anti-addiction treatments \cite{schottlender2022drugs,zhou2024artificial}.

  % ### 4. **Natural Language Processing (NLP) for Literature Mining**
  \subsection{Natural Language Processing for Literature Mining}
  NLP techniques have demonstrated convincing capabilities in processing sequential representations of biomolecules, such as SMILES and protein sequences, making them a powerful tool for target identification in drug discovery. In addiction research, NLP can also be used for text mining, which can mine valuable relationships between addiction-related targets, pathways, and drugs from a large amount of biomedical literature \cite{zheng2019text}. Models such as BERT and its variants trained on biomedical literature, such as BioBERT \cite{zhao2021recent}, can effectively identify key biomedical entities such as genes, proteins, drugs, and addiction-related pathways from scientific articles. These models also help extract relationships, allowing researchers to understand the connection between various biomedical entities, such as how a specific protein or gene is involved in the addiction process \cite{perera2020named}.

\begin{figure}[!ht]
      \centering
      \includegraphics[width=0.3\textwidth]{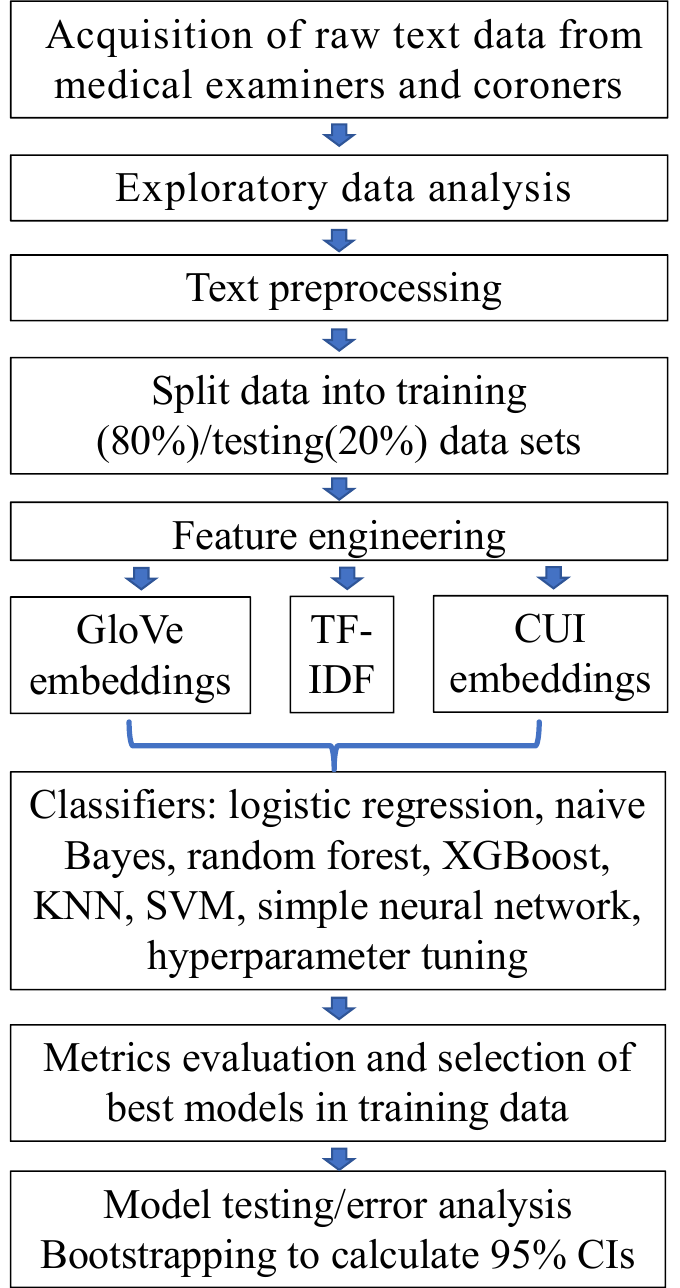}
      %\captionsetup{width=8cm}
      \caption{Natural Language Processing Pipeline for Substance Classification in Overdose Fatalities.
          Reproduced with permission from ref \cite{goodman2022development}. Copyright 2022 American Medical Association.
      }
      \label{fig:test_mining}
  \end{figure}

  In addition, NLP models such as those built on specific topics can also help organize and group addiction research articles, making it easier to identify emerging trends and related research \cite{lu2019investigate}. This technology enables researchers to efficiently process large amounts of unstructured text and discover new connections that might otherwise be overlooked, which is particularly valuable for large-scale literature mining. For example, as shown in Figure ~\ref{fig:test_mining}, \citet{goodman2022development} applied text mining techniques to classify substances associated with overdose deaths in 35,433 unstructured medical examination records in 2020 using NLP and ML. Using text mining methods, the study achieved excellent classification performance for substances such as opioids, methamphetamine, cocaine, and alcohol. NLP-based methods can also generate hypotheses about potential biomarkers or drug targets for addiction treatment, greatly accelerating the research process \cite{griffith2024natural}. Transformer models, including BERT and GPT, have revolutionized biomedical text mining by extracting therapeutic targets, summarizing findings, and highlighting promising addiction drug candidates. As a result, NLP has become an indispensable tool in addiction research, enabling faster and more comprehensive analyses, and ultimately contributing to the discovery of new therapeutic targets and biomarkers for addiction treatment.

\section{Lead Compound Discovery and Optimization}

  Lead compound discovery and optimization are essential in transforming initial compounds into effective, safe, and reliable medications for addiction treatment. The process balances scientific, clinical, and ethical considerations to address the unique challenges of addiction research, ultimately aiming to reduce the global burden of substance use disorders.
  
  \subsection{Virtual Screening}
  AI-based virtual screening for anti-addiction research represents a groundbreaking advancement in the identification of potential lead compounds from vast chemical libraries \cite{shen2023svsbi}. This innovative approach integrates computational techniques to streamline and accelerate the drug discovery process. The methodology is structured into three primary phases: high-throughput screening,   simulation, and scoring and ranking, each using the capabilities of AI\cite{li2021machine}.
  
  \paragraph{High-Throughput Screening.} 
  High-throughput screening (HTS) has become an indispensable tool in addiction-related research, especially when combined with machine learning-based techniques to construct quantitative structure-activity relationship (QSAR) models \cite{dudek2006computational,achary2020applications}. QSAR models play a key role in identifying potential anti-addiction compounds by predicting binding affinity and assessing the abuse liability of various compound classes. For example, \citet{lee2020qsar} used partial least squares regression (PLSR) to develop a QSAR model to predict the binding affinity of synthetic cannabinoids (SC) to cannabinoid receptor 1 (CB1R), a target that is closely associated with dependence and abuse potential \cite{lee2020qsar}. Similarly, QSAR models have been used to evaluate GABA-A receptor binding of emerging benzodiazepines, helping to assess their addiction and abuse potential \cite{waters2018use}. These models have also been applied to addiction-related targets such as cannabinoid CB1 and CB2 receptors, hallucinogenic phenylalkylamines, and 5-HT2A receptor binding for compounds such as phenylalkylamines, tryptamines, and LSD, as well as methcathinone selectivity for dopamine, norepinephrine, and serotonin transporters \cite{waters2018use}. Recent advances have extended QSAR modeling to more general approaches, combining advanced mathematical frameworks \cite{chen2021algebraic} and transformer models \cite{chen2021extracting}, allowing prediction of a wider range of compound properties.

  Deep learning-based models can be used to analyze vast chemical spaces and identify promising drug candidates. For example, opioids are a broad class of drugs that interact with receptors in the brain and body, such as $\mu$-opioid receptors (mORs), to relieve pain. Synthetic opioids such as fentanyl exhibit exceptionally strong binding to these receptors, and this mechanism has been extensively studied by \citet{mahinthichaichan2021kinetics}. However, the high potency and strong receptor interactions of opioids often lead to severe side effects, including addiction, so developing safer alternatives is a key area of research. To address this, \citet{szymanski2011adaptation} used AI to screen billions of virtual molecules and identified two promising compounds that have the potential to relieve pain without the addictive side effects of opioids. In addition, \citet{feng2023virtual} used machine learning to assess the hERG cardiotoxicity of compounds in the DrugBank database to aid early drug discovery. This work combines and leverages molecular sequence embedding NLP methods (autoencoders, transformers) and 3D structures with topological Laplacians and algebraic graphs through machine learning tools to develop classifiers to identify hERG blockers and regressors to analyze binding potency. The final analysis identified 227 of 8641 DrugBank compounds as potential hERG blockers, indicating safety concerns and guiding further experimental testing. \citet{chen2024machine} targets voltage-gated sodium channels (Nav1.7 and Nav1.8) to improve pain management. Using protein-protein interaction networks and machine learning, the study screened over 150,000 drug candidates for efficacy, side effects, and ADMET properties. The goal was to identify promising compounds with optimal pain treatment properties, minimizing side effects. These AI-driven methods generally outperformed traditional QSAR methods in predicting compound properties and addiction-related activities.

  In summary, the scalability and efficiency of modern HTS systems further amplify the impact of these AI models during virtual screening. HTS can enable the screening of up to 100,000 compounds per day, greatly accelerating the discovery of anti-addiction drugs \cite{szymanski2011adaptation}. Integration with advanced analytical techniques such as NMR and LC-MS/MS can fully characterize promising drug candidates, while miniaturization can reduce sample requirements and costs. Leveraging the predictive power of AI-driven QSAR models and the high-throughput capabilities of HTS, researchers can efficiently identify and develop novel anti-addiction compounds. This integrated approach represents a powerful framework for advancing addiction treatment and prevention, providing a promising path to more effectively address substance use disorders.

  \paragraph{Scoring and Docking Simulation.}
  Molecular docking is a cornerstone of structure-based drug design, which allows the prediction of the binding geometry of ligands to their target proteins through structural complementarity \cite{xu2022systematic}. This process involves systematically exploring multiple conformations of each molecule in a compound library, evaluating docking poses using scoring functions, and identifying binding modes that are consistent with experimentally observed interactions \cite{jain2006scoring,sapundzhi2019survey,li2019overview}. By repeatedly evaluating these poses, researchers focus on the most relevant ligand conformations, significantly improving the precision and effectiveness of drug discovery efforts. Among them, the construction of scoring functions is an integral part of the docking process, which usually adopts mathematical models to approximate the binding affinity between docked molecules \cite{jain2006scoring,saikia2019molecular}. These functions are crucial for estimating the strength of intermolecular interactions between ligands and their protein targets and for ranking binding poses within the target binding site \cite{quiroga2024developing}. Various scoring methods, including force field methods that exploit physics-based energy terms, empirical methods that incorporate weighted energy terms from experimental data, and knowledge-based models that exploit statistical analysis of protein-ligand complexes, have helped improve the reliability of binding predictions \cite{rajamani2007ranking}. In recent years, advanced mathematical techniques have also been incorporated into the development of scoring functions, such as topological deep learning methods \cite{cang2017topologynet}, algebraic graph-based methods \cite{nguyen2019agl}, differential geometry-based models \cite{nguyen2019dg}, and other mathematically inspired techniques \cite{chen2023path,chen2024multiscale,chen2023persistent}. To further improve the accuracy of docking and scoring, advanced computational techniques have been integrated into the workflow \cite{chao2024integration}. For example, \citet{bera2019use} leverages molecular dynamics simulations to provide flexibility to docking models, capturing the dynamics and nuances of protein-ligand interactions for a more realistic understanding of the drug binding process. \citet{quiroga2024developing} combines empirical methods with machine learning algorithms, and this hybrid approach has become a powerful and versatile tool that can be generalized to a variety of drug targets. In addition, target-specific scoring functions tailored for specific drug targets (such as proteases or protein-protein interactions) can improve prediction accuracy in specific environments\cite{guedes2021new}. Using these sophisticated docking and scoring methods, researchers can significantly simplify the drug discovery process for addiction-related targets. This approach enables more precise identification and prioritization of lead compounds, ultimately accelerating the development of effective anti-addiction therapies.
  
  \subsection{Drug repurposing}
  Drug repurposing (also known as drug repositioning) involves the identification of approved drugs for new therapeutic purposes beyond their original indications \cite{pushpakom2019drug}. This approach is particularly advantageous in anti-addiction research, as it allows for faster development and lower costs compared to developing entirely new drugs. Using the safety profiles, pharmacokinetics, and clinical data of approved drugs, repurposing offers an efficient strategy for identifying new treatments for substance use disorders. It also enables researchers to explore novel mechanisms of action for treating addiction and can target common neurobiological pathways involved in various forms of addiction \cite{zhou2021drug, du2024multiscale}. Additionally, drug repurposing fills gaps in treatment options, providing new solutions for addiction types that are currently under-treated or difficult to manage.
  
  \begin{figure*}[!ht]
      \centering
      \includegraphics[width=1\textwidth]{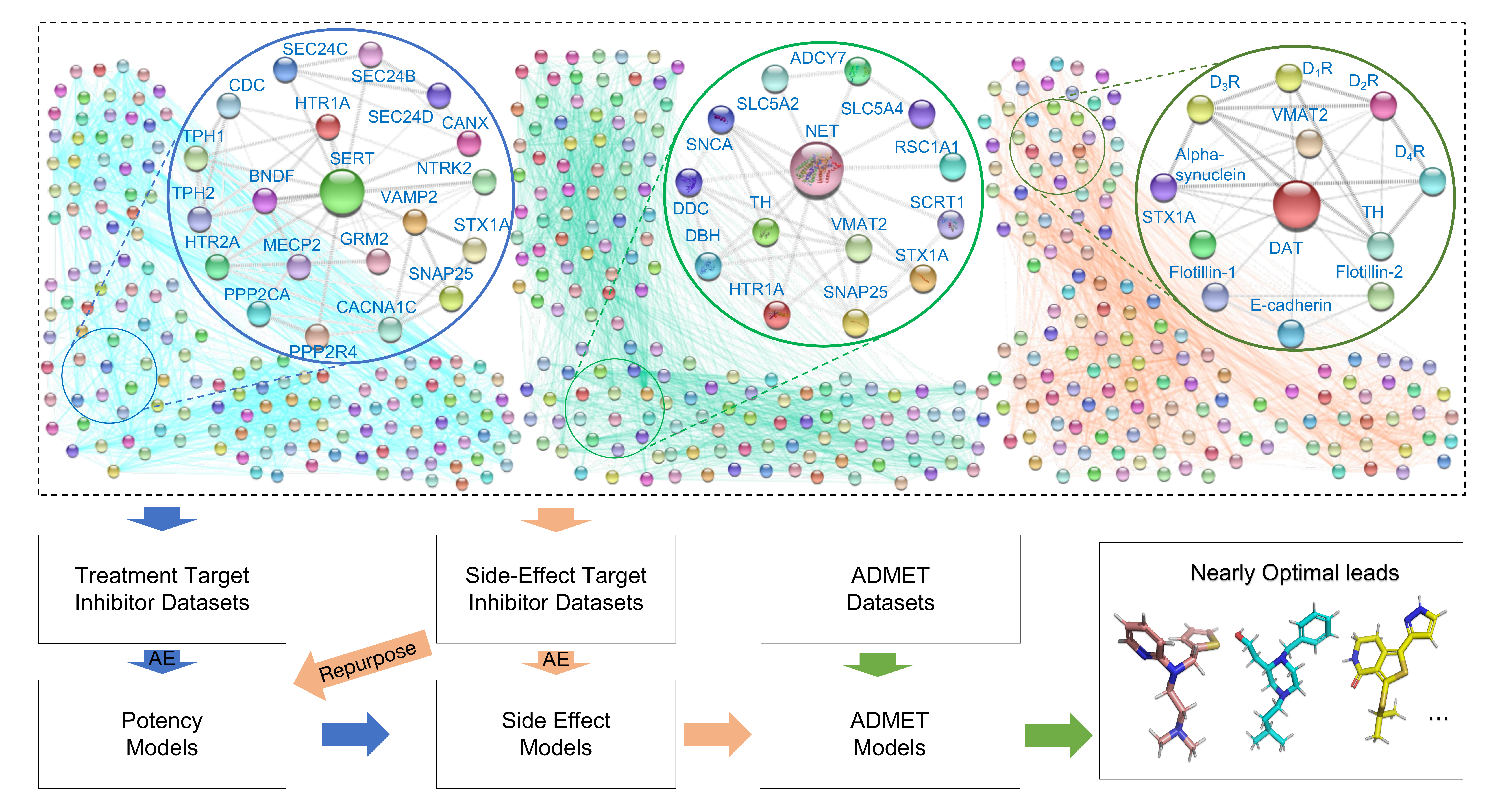}
      % \captionsetup{width=16cm}
      \caption{Core networks of DAT, SERT, and NET with a proteome-informed ML workflow for anticocaine addiction drug discovery. The figure highlights an autoencoder-based ML approach for encoding inhibitors/antagonists of proteins in DAT, SERT, and NET networks, predicting binding affinities, and identifying drug leads. Key processes include screening DAT, SERT, or NET inhibitor datasets and repurposing inhibitors/antagonists from other targets, followed by ADMET screening to refine potential leads. 
      Reproduced with permission from ref \cite{feng2022machine}. Copyright 2022 ACS Publications.
      }
      \label{fig:repurposing}
  \end{figure*}

  Advanced computational approaches have greatly improved drug repurposing strategies for addiction treatment. Network-based drug discovery systems model the phenotypic and genetic relationships among drugs, drug phenotypes, and genes, facilitating the identification of potential anti-substance use disorder candidates \cite{zhou2021drug}. As shown in Figure~\ref{fig:repurposing}, the workflow proposed by \citet{feng2022machine} uses machine learning and deep learning to address cocaine addiction, a global psychosocial disorder with no FDA-approved treatments. Focusing on dopamine (DAT), serotonin (SERT), and norepinephrine (NET) transporters, the study analyzes their protein-protein interaction networks. By examining 61 protein targets with substantial inhibitor datasets, predictive models were developed using autoencoders, GBDT, and multitask DNN to analyze 115,407 inhibitors for drug repurposing potential and side effects.
  
  Another notable advancement is the proteome-informed machine learning platform introduced by \citet{gao2021proteome}, which analyzes protein-protein interaction networks related to cocaine dependence and screens more than 60,000 drug candidates. This platform evaluates side effects, repurposing potential, and ADMET (absorption, distribution, metabolism, excretion, and toxicity) properties, ultimately identifying several promising lead compounds despite the failure of many existing drugs in the screenings. In addition, \citet{feng2023machine} applied machine learning to explore opioid receptor networks for opioid use disorder (OUD). This study screened over 120,000 drug candidates while assessing ADMET properties and identified promising inhibitors targeting nociceptin receptors. Together, these computational strategies enhance the potential for drug repurposing in addiction research by integrating diverse data sources and modeling techniques.
  
  To validate computational predictions, researchers have turned to clinical corroboration through large-scale Electronic Health Records (EHRs). By analyzing millions of patient records, researchers have evaluated the association between repurposed drugs and improved outcomes in opioid use disorder, particularly with increased rates of remission. These findings highlight the real-world applicability and potential of computationally identified drug candidates \cite{zhou2021drug}. In addition to computational and clinical validation, genetic and functional analysis has been conducted to elucidate how repurposed drugs target addiction-related pathways. These analyses focus on key molecular processes such as opioid signaling, G-protein activation, serotonin receptors, and GPCR signaling, all of which play crucial roles in addiction. Understanding how drugs interact with these pathways is essential for refining therapeutic strategies and improving the efficacy of repurposed treatments \cite{zhou2021drug}.
  
  The potential of drug repurposing in anti-addiction research is further demonstrated by its specific applications in treating various substance use disorders. For opioid use disorder, repurposed drugs such as tramadol, olanzapine, mirtazapine, bupropion, and atomoxetine have shown promise. In alcohol use disorder, drugs like topiramate, zonisamide, varenicline, and ondansetron have demonstrated potential efficacy \cite{aubin2024repurposing}. Researchers have also identified molecular targets such as mTOR, mGluR5, and NMDAR, which could provide therapeutic solutions for both opioid and cocaine addiction \cite{du2024multiscale}. These combined approaches showcase the significant promise of drug repurposing in the fight against addiction. By integrating computational predictions, clinical evidence, and mechanistic insights, drug repurposing offers a more efficient and cost-effective path to identifying new treatments for SUDs, addressing the urgent need for effective addiction therapies.
  
  \subsection{De novo drug design}
  De novo drug design, which combines generative models, reinforcement learning, and biological knowledge, can be used to create novel lead compounds or modify existing scaffolds, providing a promising approach to accelerate the discovery of anti-addiction treatments. Among them, generative models are at the core of de novo drug design, which allows the creation of new molecular structures with desired properties. These models (such as RNNs, diffusion models, VAEs, and GANs) are good at generating compounds that match the profile of addiction treatment. In particular, RNNs can identify patterns in molecular data to create new structures, while VAEs can explore chemical space and generate diverse but drug-like compounds\cite{bian2021generative,tong2021generative}. On the other hand, GANs generate highly novel and diverse molecular structures that may lead to breakthrough treatments\cite{mouchlis2021advances}. By training these models on a dataset of known effective anti-addiction compounds, molecules with the necessary properties for addiction treatment can be designed. In the de novo design process, optimizing generative models can achieve specific goals in drug design, such as anti-addiction drug design. Commonly used releases include reinforcement learning (RL), where the reward function can be defined based on key properties of successful addictive drugs, such as binding affinity to addiction-related targets (e.g., dopamine receptors), reduced abuse potential, safety, and effectiveness in controlling withdrawal symptoms\cite{wang2023chatgpt,feng2023multiobjective}.

  \begin{figure*}[!ht]
      \centering
      \includegraphics[width=0.6\textwidth]{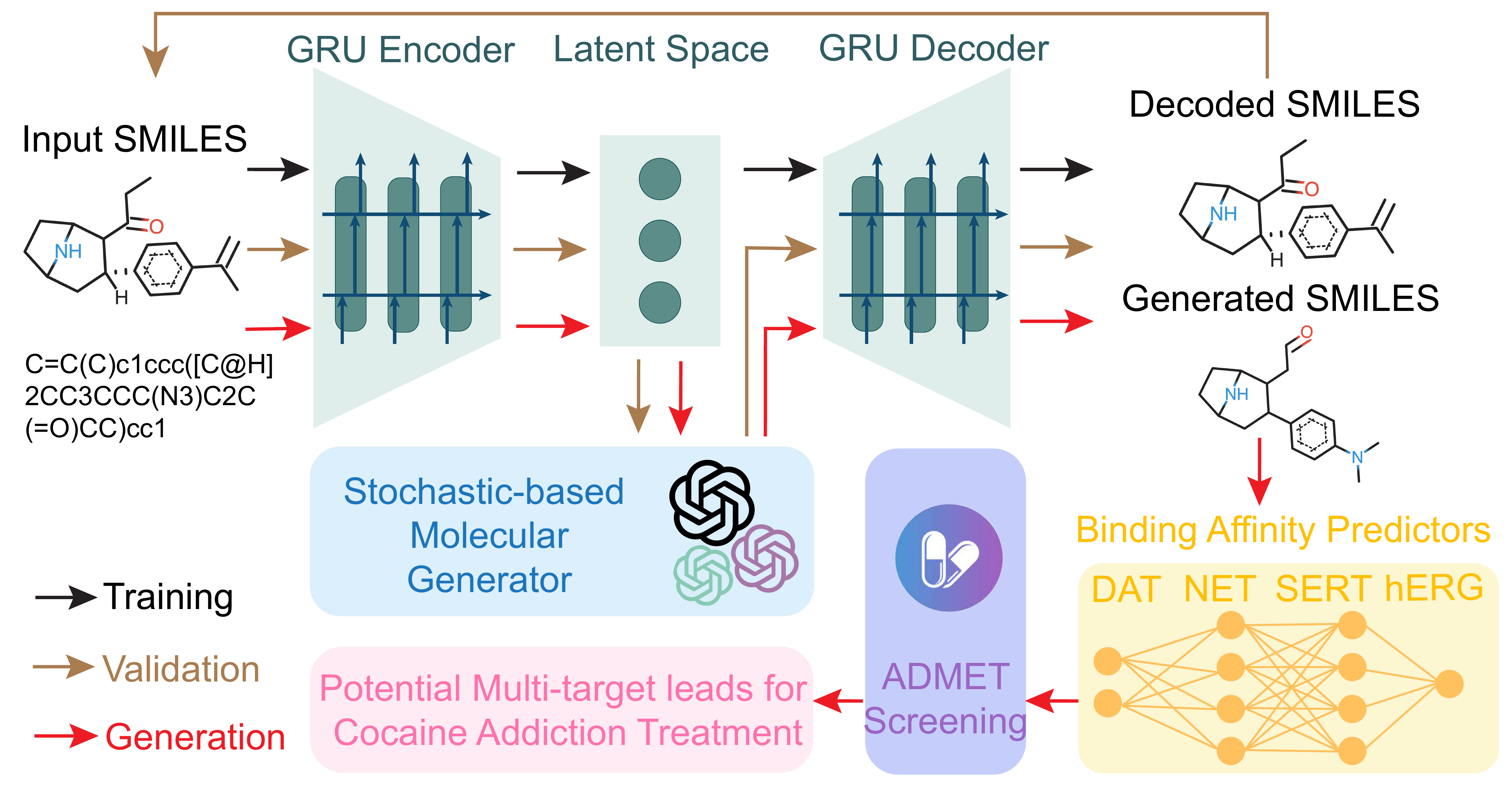}
      % \captionsetup{width=16cm}
      \caption{Overview of the Stochastic-Based Generative Network Complex (SGNC) Workflow.
      The workflow is depicted with distinct arrow colors: dark arrows represent the training process, brown arrows indicate validation and red arrows denote generation. The SGNC framework includes four main components: (1) a sequence-to-sequence AutoEncoder (green) for encoding and decoding molecular data, (2) binding affinity predictors (yellow) for assessing interactions, (3) a stochastic-based molecular generator (blue) for creating novel molecules, and (4) ADMET screening via ADMETlab 2.0 (purple) to evaluate absorption, distribution, metabolism, excretion, and toxicity properties.
      Reproduced with permission from ref \cite{wang2023chatgpt}. Copyright 2023 ACS Publications.
      }
      \label{fig:drug_design}
  \end{figure*}

  To ensure that the generated compounds are biologically relevant, integration with existing biological knowledge is critical. By incorporating biomedical knowledge graphs, the generation process can be guided to align with known addiction biological pathways and mechanisms of action. This ensures that the generated molecules interact with key addiction-related targets, such as dopamine D3 receptors or metabotropic glutamate receptor 5 (mGlu5)\cite{groman2022reinforcement}. This approach increases the likelihood that the compounds will be further developed and ultimately used clinically. By combining these computational techniques with biological insights, de novo drug design provides a faster and more efficient path to discover novel, effective, and safer addiction treatments. This integrated approach is expected to overcome challenges in addiction research and drug development.
  
  \subsection{Lead Optimization}
  Lead compound optimization is a critical stage in drug discovery, which focuses on improving candidate molecules to enhance their efficacy, safety, and drug-like properties. This step is particularly important for addiction-related research, as treatments must address the complex interactions of biological targets and pathways related to addiction. AI-driven methods play a vital role in addressing these complexities, increasing the success rate of lead compounds. Since addiction diseases involve interactions between multiple biological targets, such as dopamine, opioid, and serotonin pathways, and are further complicated by the possibility of side effects and polydrug interactions, the optimization of anti-addiction drug lead compounds presents unique challenges. To this end, multi-target optimization and multi-objective optimization were introduced to predict interactions with addiction-related targets and identify compounds with polypharmacological potential. These methods can provide insights into compounds that modulate multiple targets, ensuring a more comprehensive treatment approach.
  
  % \paragraph{Multiple objective optimizations.}
  To address the multifaceted nature of addiction treatment, multi-objective optimization plays a central role in lead compound optimization. This approach balances competing properties such as binding affinity, selectivity, drug-likeness, and ADMET (absorption, distribution, metabolism, excretion, and toxicity) properties. Techniques such as Pareto optimization \cite{yang2024enabling} allow researchers to make trade-offs between these properties, ensuring balance and enhancing therapeutic potential. For example, optimizing lead compounds often requires balancing potency and selectivity to minimize side effects while maintaining efficacy. The deep generative model developed by \citet{feng2023multiobjective,angelo2023multi} utilized the multiple-objective optimization, the model demonstrates the power of combining a stochastic differential equation-based diffusion model with a pre-trained autoencoder. This approach generates molecules targeting multiple opioid receptors (mu, kappa, delta) and rigorously evaluates their drug-likeness, ADMET characteristics, and pharmacokinetics. These models streamline the drug discovery process, ensuring that candidate compounds not only meet the therapeutic target but also satisfy critical safety and pharmacokinetic requirements for addiction treatment\cite{suva2024artificial,feng2022machine,feng2023machine}.

  % \paragraph{Multi-target optimization.}
  Multi-target optimization is also an essential strategy in addiction research, aiming to design compounds that can modulate multiple biological targets involved in the complex interactions of addiction pathways. By integrating polypharmacology and advanced computational methods, this approach ensures that the treatment is effective and well tolerated, thereby addressing the multifaceted challenges posed by addiction. In particular, polypharmacology plays a key role in identifying compounds that act on addiction-related targets such as dopamine, opioids, and glutamate receptors. The QSAR model developed by \citet{jorgensen2009efficient} uses molecular descriptors to predict the relationship between chemical structure and biological activity, which can efficiently identify compounds that are active in multiple targets. To enhance compound specificity, \citet{wang2022target} introduced a target-specific selectivity framework that decomposes selectivity into two components: 1) Absolute potency: binding affinity to the target of interest. 2) Relative potency: binding affinity compared to other targets. This bi-objective optimization framework identifies highly selective compound-target pairs, providing insights beyond traditional metrics through statistical evaluations such as permutation-based significance tests. \citet{zitnik2018modeling} proposed the Decagon model, which uses multimodal graphs to integrate various biological data, providing a powerful platform for polypharmacology modeling. These include protein-protein interactions, drug-protein target interactions, and drug-drug interactions associated with side effects. Decagon uses GCN for multi-relationship link prediction to predict specific side effects of drug combinations, leveraging extensive pharmacogenomics and patient datasets to prioritize and mitigate the risk of polypharmacy. 

Generative models further enhance multi-objective optimization by designing molecules that address multiple addiction-related pathways. POLYGON \cite{munson2024novo} is a reinforcement learning-based framework that generates novel molecular structures optimized for objectives such as polypharmacological activity, drug similarity, and synthetic feasibility. Similarly, \citet{wang2023chatgpt} highlighted the innovative application of ChatGPT as a virtual guide for anti-cocaine drug discovery, as shown in Figure~\ref{fig:drug_design}. Guided by GPT-4, a stochastic approach was integrated into the GNC model to optimize the latent space for multi-target lead generation targeting DAT, NET, and SERT using the Langevin equation. This approach integrates autoencoders, ADMET screening, and multi-target optimization, facilitating the collaboration of AI and humans to generate optimized drug candidates.

Balancing competing objectives such as binding affinity, drug-likeness, and ADMET properties is also critical for multi-target optimization. In addition, integrated platforms combining cloud computing, chemical genomics knowledge bases, and GPU-accelerated machine learning can simplify the identification and optimization of multi-target compounds. For example, for studies such as central nervous system drug treatments, these platforms are enhanced by cheminformatics tools and can be used to rapidly develop drug candidates for addiction treatment\cite{xie2014chemogenomics}. By integrating multi-target and multi-objective optimization strategies, researchers can refine lead compounds more efficiently, paving the way for breakthroughs in addiction-related therapies. These innovative computational approaches accelerate the discovery process, offering hope for the development of safer and more effective treatments for addiction disorders.

  Furthermore, side effect prediction is another key aspect of anti-addiction-related lead compound optimization. Even in complex treatment settings such as combination therapies, AI models can significantly improve this process by predicting side effects. For example, \citet{chen2022artificial} used convolutional neural networks to predict the side effects of polypharmacy. \citet{galeano2022machine} proposed a Geometric Self-Expressive Model (GSEM) to predict unknown side effects through pharmacological graph networks. \citet{lee2021toward} developed a counter-screening platform that combines machine learning-based QSAR modeling, experimental validation, and molecular simulations to design atypical DAT inhibitors for the treatment of psychostimulant use disorder. The approach ensures robust DAT binding while minimizing off-target effects, especially hERG channel binding, and identifies structural elements for optimizing lead compounds with desired pharmacological properties.

\section{Challenges and Opportunities}

  \subsection{Challenges and Limitations}
  AI-driven drug discovery for addiction faces significant challenges in the technical, biological, and ethical domains. One of the main hurdles is the scarcity of labeled data specific to addiction-related drug targets, which AI models, particularly deep learning approaches, require to train effectively\cite{blanco2023role,dou2023machine}. This data shortage, combined with scalability issues, limits the application of traditional machine learning tools in handling the large datasets typical in drug discovery\cite{mak2024artificial}.
  
  The complexity of addiction mechanisms further complicates the process. Addiction involves intricate interactions between neurotransmitter systems such as the dopamine, opioid, and glutamate pathways, making it difficult for AI models to predict drug efficacy accurately\cite{wang2023chatgpt}. This challenge is amplified by the lack of clear biomarkers to access treatment outcomes, hindering the validation of AI-driven approaches\cite{mak2024artificial}.
  
  Ethical concerns are also significant, particularly concerning the interpretability of AI models. Their "black box" nature makes it challenging to trace decision-making processes, raising issues of transparency\cite{wu2023black}. Furthermore, biases in training data can lead to skewed predictions, emphasizing the need for fairness and representativeness in AI-based drug development\cite{blanco2023role}.
  
  On the regulatory front, the rapid pace of AI advancements has created a regulatory gray area, with bodies like the FDA working to establish appropriate frameworks \cite{gottlieb2023regulators}. However, the opacity of AI algorithms complicates the traceability of decisions, particularly in clinical trials \cite{gottlieb2023regulators}. The continuous learning nature of AI models also challenges traditional regulatory models, which evaluate fixed products. New assessment methods may be needed, such as validating AI predictions against historical data. Furthermore, while AI can accelerate drug discovery, rigorous experimental validation is still essential to ensure real-world applicability\cite{visan2024integrating}. An iterative feedback loop between AI predictions and experimental data is necessary to verify the safety and efficacy of drug candidates. Addressing these regulatory and validation challenges is crucial to ensure the successful application of AI in addiction treatment.
  
  \subsection{Trends and Directions} 
  Despite these challenges, recent advances in AI and related technologies are driving transformative progress in addiction-related drug discovery. AI-powered structure-based drug discovery is enabling researchers to analyze molecular interactions with unprecedented precision by combining machine learning with physics-based simulations\cite{choudhury2022structure}. This integration facilitates the design of highly targeted and effective therapeutic compounds.
  
  AI-driven virtual screening represents a highly promising avenue in drug discovery \cite{shen2023svsbi}. By leveraging advanced NLP models to generate embeddings for both targets and drugs, this approach eliminates the need for costly molecular docking procedures. Moreover, it enables automated, large-scale virtual screening across multiple databases, significantly enhancing efficiency and scalability.
  
  To overcome the challenge of data scarcity, transfer learning or multitask learning has emerged as a valuable approach\cite{cang2017topologynet}. By leveraging knowledge from broader datasets, AI models can adapt insights to the specialized field of addiction research, improving efficiency and accuracy. Federated learning also addresses data limitations by enabling collaborative model development without requiring organizations to share sensitive data, ensuring privacy and ethical compliance\cite{chen2020fl,kumar2021federated}.  Moreover, generative AI is another popular approach to overcome data scarcity \cite{gao2020generative,dou2023machine}. 
  
  Another promising trend is the integration of AI with high-throughput experimental techniques, such as cryo-electron microscopy (cryo-EM)\cite{danev2019cryo} and high-content screening \cite{giuliano2003advances}. This synergy accelerates the discovery and validation of drug candidates, significantly shortening the path to clinical application. In addition, AI-driven precision medicine approaches are revolutionizing addiction treatment by tailoring therapies to individual patient characteristics, such as genetic profiles and medication responses. This personalized approach improves treatment efficacy and reduces relapse rates\cite{blanco2023role}.
  
  The integration of (multi-)omics data and advanced data analysis techniques is rapidly emerging as a transformative approach to anti-addiction drug discovery \cite{du2024multiscale}. In particular, spatial transcriptomic analysis facilitates precise target identification and provides an efficient means to evaluate the effectiveness of anti-addiction drugs.
  
  The application of large language models (LLMs) in anti-addiction drug discovery is expected to become a prominent topic of interest \cite{wang2023chatgpt}. Using their capacity to analyze vast datasets, handle complex information, and uncover insights that were previously challenging to obtain, LLMs hold significant promise for transforming the drug discovery process.
  
  Mathematical deep learning (MathDL) and topological deep learning (TDL) have achieved remarkable success in drug design \cite{cang2017topologynet}. These methodologies emerged as top performers in the D3R Grand Challenges, a global competition series focused on advancing computer-aided drug design \cite{nguyen2019mathematical,nguyen2020mathdl}. The continued development of mathematical AI promises to drive innovation and create transformative methods in drug discovery.   
  
  Together, these advancements highlight AI's potential to overcome existing barriers and provide innovative solutions for addiction treatment. Continued interdisciplinary collaboration among AI experts, addiction researchers, and healthcare professionals will be crucial to address the multifaceted challenges of this field and advance the development of effective patient-centric therapies.

\section*{Author contributions}
  Chen collected references and drafted the initial manuscript. Jiang and Su revised the manuscript.  Wei supervised the project and revised the manuscript. All authors approved the final version of the manuscript.
  
\section*{Conflicts of interest}
  The authors declare that they have no competing interests.
  
\section*{Data availability Statement}
This review does not include any new data. All data referenced are available in the cited sources.

\section*{Acknowledgments}
This work was supported in part by National Institutes of Health grants R01GM126189, R01AI164266, and R35GM148196, National Science Foundation grants DMS2052983 and IIS-1900473, Michigan State University Research Foundation, and  Bristol-Myers Squibb 65109.

%%%END OF MAIN TEXT%%%


\balance


%%%REFERENCES%%%
\bibliography{main_arxiv} %You need to replace "rsc" on this line with the name of your .bib file
\bibliographystyle{main_arxiv} %the RSC's .bst file
\end{document}